\documentclass[10pt,letterpaper]{article}
\usepackage[top=0.85in,left=2.75in,footskip=0.75in,marginparwidth=2in]{geometry}
\usepackage{textcomp}

\usepackage[utf8]{inputenc}

\usepackage{cite}
\usepackage[T1]{fontenc}
\usepackage{indentfirst}
\usepackage{amssymb}
\usepackage{amsmath}
\usepackage{nameref,hyperref}

\usepackage[right]{lineno}

\usepackage{microtype}
\DisableLigatures[f]{encoding = *, family = * }

\raggedright
\usepackage{changepage}
\usepackage{ragged2e}
\usepackage[aboveskip=1pt,labelfont=bf,labelsep=period,singlelinecheck=off]{caption}

\makeatletter
\renewcommand{\@biblabel}[1]{\quad#1.}
\makeatother

\usepackage{lastpage,fancyhdr,graphicx}
\usepackage{epstopdf}
\fancyheadoffset[L]{2.25in}
\fancyfootoffset[L]{2.25in}

\usepackage{color}

\definecolor{Gray}{gray}{.25}

\usepackage{graphicx}

\usepackage{sidecap}

\usepackage{wrapfig}
\usepackage[pscoord]{eso-pic}
\usepackage[fulladjust]{marginnote}
\reversemarginpar

\begin{document}
\vspace*{0.35in}

\begin{flushleft}
{\Large
\textbf\newline{CPSC: Conformal prediction with shrunken centroids for efficient prediction reliability quantification and data augmentation, a case in alternative herbal medicine classification with electronic nose.}
}
\newline
\\
Li Liu\textsuperscript{1},
Xianghao Zhan\textsuperscript{1},
Xikai Yang\textsuperscript{2},
Xiaoqing Guan\textsuperscript{1},
Rumeng Wu\textsuperscript{1},
Zhan Wang\textsuperscript{1},
Zhiyuan Luo\textsuperscript{3},
You Wang\textsuperscript{1,*},
Guang Li\textsuperscript{1}
\\
\bigskip
\bf{1} The State Key Laboratory of Industrial Control Technology, Institute of Cyber-Systems and Control, Zhejiang University, Hangzhou 310027, China.
\\
\bf{2} University of Michigan - Shanghai Jiao Tong University Joint Institute, Shanghai Jiao Tong University,China
\\
\bf{3} The Computer Learning Research Center, Royal Holloway, University of London, Egham Hill, Egham, Surrey TW20 0EX, UK
\\

\bigskip
* king\_wy@zju.edu.cn

\end{flushleft}

\section{Abstract}
\justifying
In machine learning applications, the reliability of predictions is significant for assisted decision and risk control. As an effective framework to quantify the prediction reliability, conformal prediction (CP) was developed with the CPKNN (CP with kNN). However, the conventional CPKNN suffers from high variance and bias and long computational time as the feature dimensionality increases. To address these limitations, a new CP framework-conformal prediction with shrunken centroids (CPSC) is proposed. It regularizes the class centroids to attenuate the irrelevant features and shrink the sample space for predictions and reliability quantification. To compare CPKNN and CPSC, we employed them in the classification of 12 categories of alternative herbal medicine with electronic nose as a case and assessed them in two tasks: 1) offline prediction: the training set was fixed and the accuracy on the testing set was evaluated; 2) online prediction with data augmentation: they filtered unlabeled data to augment the training data based on the prediction reliability and the final accuracy of testing set was compared. The result shows that CPSC significantly outperformed CPKNN in both two tasks: 1) CPSC reached a significantly higher accuracy with lower computation cost, and with the same credibility output, CPSC generally achieves a higher accuracy; 2)  the data augmentation process with CPSC robustly manifested a statistically significant improvement  in prediction accuracy with different reliability thresholds, and the augmented data were more balanced in classes. This novel CPSC provides higher prediction accuracy and better reliability quantification, which can be a reliable assistance in decision support.

\section{Keywords}
conformal prediction, shrunken centroids, reliability, herbal medicine, electronic nose

\section{Introduction}
\label{Introduction}
Machine learning classification algorithms have enabled many successful applications in the field of biomedicine such as patient outcome prediction \cite{covid19} and free-text clinical note mining \cite{clinic_text}. For classification applications in biomedicine, besides the metrics like accuracy, precision and recall, the reliability of the predictions is of great significance, as the misclassification can lead to delayed treatment and tremendous physiological and mental burden \cite{2cpknn_lung_cancer}. For instance, in the case of cancer diagnosis, the risky prediction with unclear reliability can either lead to undetected carcinoma and even metastasis, or the false positives which may lead the patient to suffer from costly and harmful chemotherapy and radiotherapy. On the contrary, if the reliability of predictions could be quantified, additional and complementary diagnostic approaches can be taken in time to improve the diagnosis reliability.\\

\indent 
To quantify the prediction reliability in machine learning, Platts method \cite{platts_method}, probably approximately correct learning (PAC) \cite{PAC}, Bayesian learning \cite{bayesian_learning_probability}, have been proposed. However, those methods typically require the data to be in large quantities or under normal distribution to estimate a prior probability. These assumptions are hard to satistfy in real-world applications due to factors like measurement limitations, such as inadequate training samples, measurement noises, sensor drifts. Therefore, a more robust and precise reliability quantification method without strong assumptions is in need.
Conformal prediction (CP), proposed by Shafer and Vovk \cite{1cp_introduction}, has proved a computational framework effective in prediction reliability quantification. It is featured by generating accurate confidence levels of predictions: with a given significance level $\varepsilon$, it outputs a set of region predictions which contains the true label with confidence at least 1-$\varepsilon$. \\
\indent
Without the requirements of distributional or model assumptions , CP can work with any underlying algorithm to calculate nonconformity measurement \cite{cp_intro_new}, which is used to measure the conformity of a prediction with the training data. The k-nearest neighbor (kNN) was initially proposed as an underlying algorithm. In previous research, CP is frequently combined with kNN \cite{2cpknn_lung_cancer,3feature_engineering,4augmentation,cp_intro_ICP,cp_intro_KNN}, and the approach is usually referred to as CPKNN, which quantifies the nonconformity of a new sample via the ratios of the sums of Euclidean distances from k nearest observations with same/different potential labels. However, multiple issues arise when CPKNN is used:  firstly, when dealing with high-dimensional data, CPKNN suffers from the curse of dimensionality: the sparsely scattered points in high-dimensional space may lead the pairs of samples close in Euclidean distance to be not similar in their classes, particularly when many dimensions do not contain relevant classification information, which may finally lead to high variance and high bias.\cite{the_element_of_stat_learning}. Secondly, to get the k nearest observations, CPKNN needs to iteratively compute the distances on the entire training set. Therefore, CPKNN is computationally expensive. Compared with the kNN, shrunken centroid (SC) classifier, as a regularized version of the nearest centroid (NC) classifier and with the additional attenuation of noisy and uninformative dimensions, has been shown effective in fast computation and in dealing with high-dimensional data. Proposed by Tibshirani et al.\cite{5shrunken_centroid}, SC simplifies the distance computation process by taking class centroids as the reference points rather than all the scattered sample points. Therefore, SC only needs to compute the Euclidean distances with the time in an order of the number of classes rather than the number of training samples. Additionally, SC can make the high-dimensional feature space more compact by shrinking the class centroids and attenuating the irrelevant features for classification, which is meaningful to the real-world application of measurement technology, as higher dimensions of features are frequently observed in machine learning modeling, particularly in the field of biomedicine (e.g., the high-dimensional gene data and biomedical sensor measurements). Considering the prediction reliability of the high-dimensional data, it is worthwhile to develop and investigate conformal prediction based on shrunken centroids (CPSC).\\
\indent
As a appropriate case with the issue of high dimensionality in the field of biomedical research and the significance of reliability quantification, the classification of alternative herbal medicine with the electronic nose is worthwhile to be investigated \cite{e_nose_herbal_medicine,augmentation}. Different categories herbal medicines are valuable in medical research and patient treatment, but even the categories close in species can bring about diverse treatment effect \cite{herbal_medicine}. The different treatment values and varying prices of herbal medicines lead to the fraud and inferior herbal medicine products. These facts call for an effective way to identify different herbal medicines for better patient treatment. However, many of the herbal medicines are hard to classify due to their similar appearance, especially when they are pulverized. Electronic nose system (e-nose), has proved its effectiveness in solving this task. However, in previous publications, high-dimensional engineered features are extracted from e-nose signals, contributing to the potential curse of dimensionality \cite{augmentation,2-feature_engineering,3feature_engineering,E_nose_Feature_engineering,curse_of_dimensionality}. It can not only bring high variance and bias in getting accurate and reliable prediction, but also high computational cost before we get the predictions and reliability quantification with the conventional CPKNN. Therefore, it is meaningful to develop an efficient mechanism with e-nose system to further 
optimize the prediction reliabiltiy quantification and computational speed.\\
\indent In this study, a new CP framework, CPSC, is developed, and its effectiveness is compared with the conventional CPKNN on the classification task of 12 categories of herbal medicines, where both computational cost and prediction reliability were compared in two scenarios: 1) offline prediction, where the training data were fixed and three metrics on testing set were evaluated: the runtime, the prediction accuracy and the quantified reliability, 2) online prediction with data augmentation, where the training data were augmented with samples filtered by the CP in an on-going manner, and the prediction dynamics of both CP frameworks and their respective influences on the classifier's accuracy were compared. As a result, this study reveals that CPSC contributes to better reliability quantification, higher classification accuracy, more effective data augmentation and an remarkable advantage of computational speed.

\section{Method}
\subsection{Conformal Prediction}
In machine learning classification problems, the classifiers leverage the knowledge from an observed training set $Z$:$\left(\left(x_{1}, y_{1}\right), \ldots,\left(x_{n-1}, y_{n-1}\right)\right)$ to learn the mapping of a new sample $x_{n}$ from the sample space $X$ (also known as feature space) to its label $y_{n}$ in the label space $Y$ \cite{1cp_introduction,cp_intro_ICP}. With an additional parameter, the significance level $\varepsilon$, conformal predictor can give the new sample $x_{n}$ a prediction region $\Gamma^{\varepsilon}$ which is a subset of label space $Y$, with a confidence level $1-\varepsilon$. This confidence level indicates how much confidence one has for that the true label is included in the prediction region $\Gamma^{\varepsilon}$ \cite{cp_intro_KNN,Neural_engineering_ICP,cp_intro}. With this quantified probability, conformal prediction enables the users to make reasonable decisions with better risk control.

\begin{equation}
\mathbb{P}\left({y}_{n} \in \Gamma^{\varepsilon}\right) \textgreater 1-\varepsilon
\end{equation}

\indent In the process of modeling prediction regions, conformal predictors employ a mechanism called nonconformity measurement to evaluate how conformed or strange a sample is when compared with other observations already seen. For each example in $z_{i}(i=1,2, \ldots, n) \in Z$, the nonconformity measurement presents as a real number $\alpha_{i}(i=1,2, \ldots, n) \in R$, which is calculated by a measurable function $A$:
 \begin{equation}
    \label{nonconformity measurement}
   \alpha_{i}=A\left(z_{1}, \ldots z_{i-1}, z_{i+1}, \ldots, z_{n}\right), i=1, \ldots, n
    \end{equation}
\indent For a new sample $x_{n}$, P-values are computed based on the nonconformity measurement, which then enables the construction of the prediction region according to the following formula:

 \begin{equation}
\Gamma^{\epsilon}\left(\left(z_{1}, z_{2}, \ldots, z_{n-2}, z_{n-1}\right), x_{n}\right)=\left\{y \mid p^{y}>\epsilon\right\}
    \end{equation}
where $p^{y}$ is the P-value of a potential label $y$ ($y \in Y$) for the new sample $x_{n}$. It indicates how well this new sample (with the predicted label of $y$) conforms to observations already seen. With every potential label $y$ for $x_{n}$, the nonconformity measurement $\alpha_{n}^{y}$ is computed, and corresponding $p^{y}$ is defined as:
\begin{equation}
\label{pvalue counting}
    p^{y}=\frac{\left|\left\{i=1, \ldots, n \mid \alpha_{i}^{y} \geq \alpha_{n}^{y}\right\}\right|}{n}
\end{equation}

What must be mentioned is that unlike the predicted probability with a softmax function in the multi-class classification, the sum of P-values for one sample over the label space does not necessarily equal 1 because the conformal predictor does not directly model the conditional probability $P(Y|X)$, but focuses on the conformity with the set of training observations. Therefore, the nonconformity measurement of each possible label in the label space is calculated as an independent attempt based on the training observations. If CP was forced to output one prediction $y$ based on the largest P-value (also referred to as forced prediction), two reliability metrics for the prediction can be given: 1) credibility: the largest P-value, which indicates how reliable the best choice of prediction $y$ in the label space $Y$ is, based on its conformity to the training observations; 2) confidence: 1 - the second largest P-value, which indicates how reliable the best prediction $y$ is, based on the exclusion of other possible choice of prediction.

\subsection{Shrunken centroids}
As a regularized version of the nearest centroid classifier \cite{nearest_centroids}, shrunken centroid (SC) classifier attenuates the irrelevant features and shrink the class centroids to the overall centroid \cite{5shrunken_centroid}, which alleviates the negative influences on classification exerted by noises. For a test sample $x$, SC finds its nearest 'denoised' centroid based on the Euclidean distance and outputs the predicted label based on the label associated with the centroid. In this process, the new sample only needs to be compared with $K$ class centroids, instead of being compared with all the individual observations as what is done in kNN classifier. Furthermore, SC shrinks the reference sample space covered by the training samples to be more compact as a form of regularization, and therefore alleviates the curse of dimensionality intrinsic in many Euclidean-distance-based methods such as KNN. To model the posterior probabilities of a potential predicted label $k$: $P(Y=k|X)$, SC works with observed training set $Z$ in following steps: \\
\indent 1) Compute the prototype centroids $\bar{x}_{k}$ of each class (1,2,...,$K$) and the overall centroid $\mu$ of all observations  ($Z$:$\left(\left(x_{1}, y_{1}\right), \ldots,\left(x_{n}, y_{n}\right)\right)$) as (\ref{class centroid}),(\ref{overall centroid}) shows, where ${C}_{k}$ denotes the set of samples with label $k$, and ${n}_{k}$ denotes the quantity of samples with label $k$. 
\begin{equation}
\label{class centroid}
    \bar{x}_{k}=\sum_{i=1}^{C_{k}} \frac{x_{i k}}{n_{k}}
\end{equation}
\begin{equation}
\label{overall centroid}
\mu=\sum_{i=1}^{n} \frac{x_{i}}{n}
\end{equation}
\indent
2) Calculate the pooled within-class standard deviation, and standardize the bias between class centroids and the overall centroid:
\begin{equation}
\label{eq:std_update}
s^{2}=\frac{1}{n-k} \sum_{k} \sum_{i \in C k}\left(x_{i}-\bar{x}_{k}\right)^{2}
\end{equation}
\begin{equation}
d_{k}=\left(\bar{x}_{k}-\mu\right) / s
\end{equation}

3) Shrink the biases with a threshold $\Delta$:
\begin{equation}
\label{shrink_bias}
\begin{array}{c}
d_{k}^{\prime}=\operatorname{sign}\left(\mathrm{d}_{k}\right)\left(\left|\mathrm{d}_{k}\right|-\Delta\right)_{+} \\t_{+} \quad=\left\{\begin{array}{ll}t & t>0 \\0 & t \leq 0
\end{array}\right.
\end{array}
\end{equation}
\indent Note that the centroids $\bar{x}_{k}$, $\mu$, the standard deviation $s$, and biases ${d}_{k}$ are all vectors, which contain the information of all $p$ features $j$=1,2,...,$p$. As \ref{shrink_bias} shows, on the one hand, if ${d}_{{j}{k}}$, the bias of a $j$-th feature of class $k$, has an absolute value which is smaller than the threshold $\Delta$, the corresponding feature will be considered not informative enough for classification. Therefore, this bias in this feature will be shrunken to zero and the noisy feature is therefore removed, which also lowers the data dimensionality. On the other hand, if ${d}_{{j}{k}}$ has an absolute value larger than the threshold $\Delta$, the bias will be shrunken towards the overall means, shrinking the sample space as a form of regularization. \\
\indent 
4) Update the class centroids with the shrunken biases:
\begin{equation}
\bar{x}_{k}^{\prime}=\bar{x}_{k}+s*d_{k}^{\prime}
\end{equation}
\indent
5) Compare the new sample with shrunken centroids:
    \begin{equation}
    \label{eq:new_cluster}
\delta_{k}\left(x^{*}\right)=\log \pi_{k}-\frac{1}{2} \sum_{j=1}^{\mathrm{p}} \frac{\left(x_{j}^{*}-\bar{x}_{j k}^{\prime}\right)^{2}}{s_{j}^{2}}
\end{equation}
\indent This discrimination score $\delta_{k}\left(x^{*}\right)$ measures how close the new sample is to the shrunken centroids. Its result consists of two terms: the first term $\log \pi_{k}$ is the prior probability of class $k$, defined by the portion of the samples in class $k$ in overall observations. The second term is the standardized squared distance between $k$-th centroids and this new sample. Therefore, the posterior probability of a given class $k$, $P(Y=k|X)$ is decided by both the prior class distribution and sample's proximity to different centroids.\\ 
\indent 6) Compute the probability of that a new sample belongs to class $k$:
\begin{equation}
\hat{p}_{k}\left(x^{*}\right)=\frac{e^{\delta_{k}\left(x^{*}\right)/T}}{\sum_{\ell=1}^{K} e^{\delta_{\ell}\left(x^{*}\right)/T}}
\end{equation}
\indent To convert the probabilities into the range [0,1] and aggregates to 1, the softmax method is adopted. Additionally, a scale parameter $T$ was innovatively added, which may shrink the value of $\delta_{k}(x^*)$ and make the predicted probability distribution softer. We borrowed the idea of temperature from knowledge distillation in this design because the softmax function with a higher temperature generally leads to a more balanced distribution across different labels and therefore maintain the information of those less possible labels as well as avoid overfitting to some extent\cite{knowledge_distillation_1,knowledge_distillation_2}.

\subsection{Conformal prediction with Shrunken Centroids (CPSC)}
Conformal predictor leverages a measurable function A (\ref{nonconformity measurement}) as the nonconformity measurement. For SC, inspired by the previous works \cite{Neural_engineering_ICP,Neuro_computing}, the function A was defined as:
\begin{equation}
\label{A measurement}
\alpha_{i}=0.5-\frac{\hat{p}\left(y_{i} \mid x_{i}\right)-\max \hat{p}_{\mathrm{y} !=y_{i}}\left(y \mid x_{i}\right)}{2}
\end{equation}
where $y_{i}$ denotes the label for $x_{i}$. For a sample among the training observations $Z$, the nonconformity measurement is a real number because its label is known. However, for a new sample $x^{*}$, its $\boldsymbol{\alpha^{*}}$ is a vector with a length of $K$ which is the number of potential labels. This vector contains all the $\alpha_{k}^{*}$ for all the potential labels $y_{k}=1,2,...,K$ in $Y$.\\
\indent
With the nonconformity measurement $\boldsymbol{\alpha^{*}}$, the P-value $p^{*}$ of $x^{*}$ is given as (\ref{pvalue counting}) shows. Note that $\boldsymbol{p^{*}}$ is also a vector, including the P-values for all potential labels of $x^{*}$. The labels with P-values higher than the significance level $\varepsilon$ will be output in the prediction region, and the largest P-value indicates the most reliable label based on the nonconformity measurement.

\subsection{Conformal prediction with kNN (CPKNN)}
\indent In CPKNN, the function A for nonconformity measurement of $x_{i}$ is defined as \cite{E_nose_WZ_cpknn}:
\begin{equation}
\label{Nonconformity measurement CPKNN}
\alpha_{y}=\frac{\sum_{m=1}^{k_{num}} d\left(x_{i}, x_{m}^{{y}}\right)}{\sum_{\ell=1}^{k_{num}} d\left(x_{i}, x_{\ell}^{! {y}}\right)}
\end{equation}
\indent To compute it, firstly, the distances between $x_{i}$ and other observations in $X(x_{1},\ldots,x_{i-1},x_{i+1},\ldots,x_{n})$ are calculated, denoted as $d\left(x_{i}, x_{j}\right)$. Secondly, the $k_{num}$ nearest observations which have the same/different labels with $x_{i}$ are ranked respectively. Here $x_{m}^{{y}}$ denotes the observations with the same label as $x_{i}$, and the $x_{\ell}^{!{y}}$ denotes the observations with a different label. Then, the sums of the distances from $k_{num}$ nearest observations with same/different label are divided. Under this scheme, the closer a sample locates to its homogeneous groups and farther to its heterogeneous groups, the lower nonconformity measurement it gets. The way CPKNN converts $\alpha$ to P-value follows (\ref{pvalue counting}), with the same rule as CPSC to give prediction outputs based on P-value.  
\subsection{Experiment and dataset}
\indent In this study, CPSC and CPKNN were applied to the classification of 12 categories of alternative herbal medicines based on a self-assembled e-nose system. This e-nose system has been shown effective in the classification of herbal medicines in our previous publications \cite{E_nose_zhan2018sensors,E_nose_Feature_engineering,E-nose_mechanism, E_nose_ginseng,Diao_dendrobium}. The system includes 16 TGS (Taguchi Gas Sensors) type metal-oxide semi-conductive (MOS) sensors by Figaro Engineering Inc, Osaka, Japan. The details of the sensors are listed in Table \ref{Sensors_speciality}. 

\begin{figure}
    \centering
    \includegraphics[width=\linewidth]{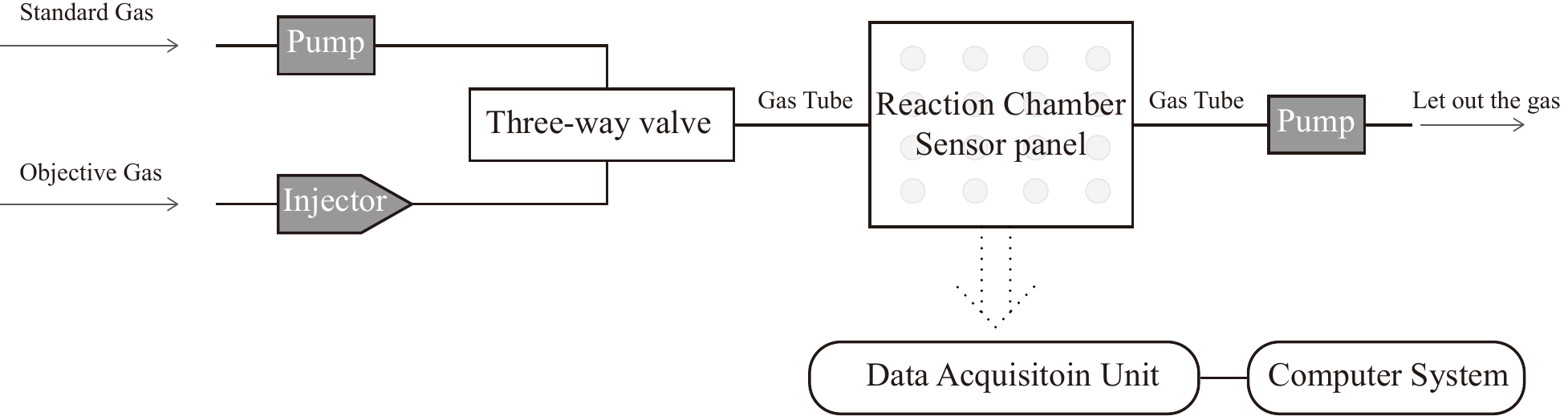}
    \caption{The brief description of the electronic nose system. The e-nose system consists of four sub-systems: 1) the gas transporting system to control the flow, 2) the reaction chamber where the objective gas reacts with the sensor panel, 3) the data acquisition unit to record the signals, and 4) the computer system for signal processing and classification.}
    \label{Fig_e_nose_system}
\end{figure}

\begin{figure}
    \centering
    \includegraphics[width=\linewidth]{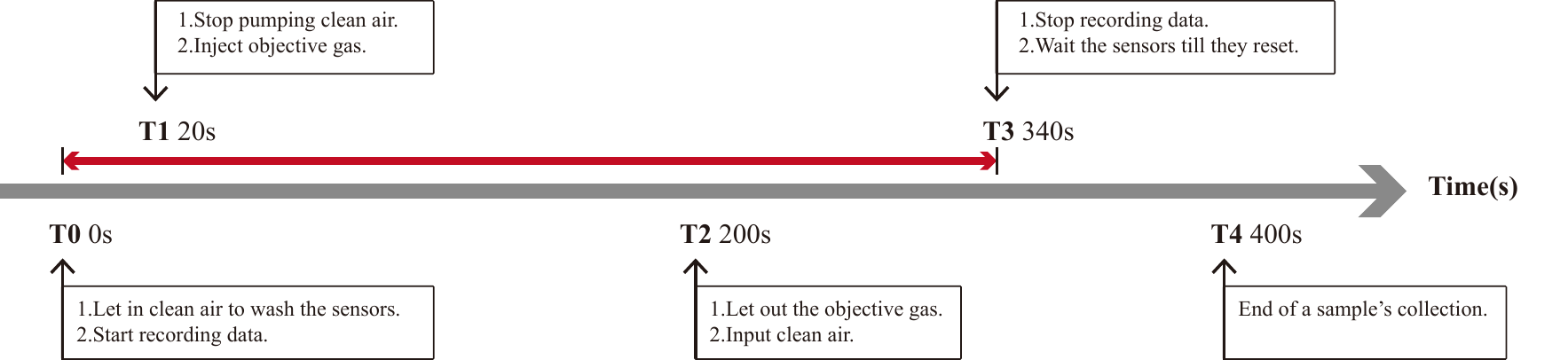}
    \caption{The description of the data collection process for an individual sample. For each sample, the experiment process took 400s, and the data recording period (denoted by the red line) lasted for 340s. The experiment contained 4 steps: 1) T0-T1: the sensors get initialized using clean air, with the baseline values recorded. 2) T1-T2: the reaction of objective gas and the sensor panel. 3) T2-T3: the outflow of the objective gas and the cleaning of the reaction chamber. 4) T3-T4: the sensors get reset.  
    }
    \label{Data_collection}
\end{figure}

\indent This self-assembled e-nose system consisted of four major components: the gas transporting system, the sensors reaction chamber, the data acquisition unit, and the data processing and pattern recognition algorithms. The brief description of e-nose system is illustrated in Fig. \ref{Fig_e_nose_system}. The gas transporting system controls the flow of the standard gas (dry and clean air) and the objective gas with two gas pumps and a three-way valve. The entire process of collecting a sample lasted for 400 seconds, and the sensors' reactions are recorded by data acquisition unit (DAQ) with a sampling rate of 100 Hz, as is shown in Fig. \ref{Data_collection}. During the intervals between the sensor reaction time periods for two different samples, the standard gas was pumped in to reset the sensors. This baseline-setting period took up 20 s with a flow rate of 1 L/minutes. The objective gas, which was the 10 mL headspace air collected from each individual sample, was then injected into the sensors reaction chamber to react with the sensor panel (shown as ascending curve in the sensor reaction signals). The objective gas stayed in the chamber for 180 s, and then it was pumped out with the input of the standard gas again (shown as descending curve in the sensor reaction signals).     

\begin{table}[]
\caption{The particular high-sensitivity volatile organic compounds for different sensors used in the study.}
\label{Sensors_speciality}
\scalebox{0.8}{
\begin{tabular}{ccl}
\hline
No. & Sensor Type & Specific Response Sensitivity                                                                                       \\ \hline
1   & TGS800      & Carbon monoxide, ethanol, methane, hydrogen, ammonia                                                                \\
2   & TGS813      & Carbon monoxide, ethanol, methane, hydrogen, isobutane                                                              \\
3   & TGS813      & Carbon monoxide, ethanol, methane, hydrogen, isobutane                                                              \\
4   & TGS816      & Carbon monoxide, ethanol, methane, hydrogen, isobutane                                                              \\
5   & TGS821      & Carbon monoxide, ethanol, methane, hydrogen                                                                         \\
6   & TGS822      & \begin{tabular}[c]{@{}l@{}}Carbon monoxide, ethanol, methane, acetone, n-hexane,\\ benzene, isobutane\end{tabular}  \\
7   & TGS822      & \begin{tabular}[c]{@{}l@{}}Carbon monoxide, ethanol, methane, acetone, n-Hexane,\\  benzene, isobutane\end{tabular} \\
8   & TGS826      & Ammonia, trimethyl amine                                                                                            \\
9   & TGS830      & Ethanol, R-12, R-11, R-22, R-113                                                                                    \\
10  & TGS832      & R-134a, R-12 and R-22, ethanol                                                                                      \\
11  & TGS880      & Carbon monoxide, ethanol, methane, hydrogen, isobutane                                                              \\
12  & TGS2620     & Methane, Carbon monoxide, isobutane, hydrogen                                                                       \\
13  & TGS2600     & Carbon monoxide, hydrogen                                                                                           \\
14  & TGS2602     & Hydrogen, ammonia ethanol, hydrogen sulfide, toluene                                                                \\
15  & TGS2610     & Ethanol, hydrogen, methane, isobutane/propane                                                                       \\
16  & TGS2611     & Ethanol, hydrogen, isobutane, methane                                                                               \\ \hline
\end{tabular}}
\end{table}

\indent The experiment was done at the State Key Laboratory of Industrial Control Technology, Zhejiang University in 2017 \cite{E_nose_zhan2018sensors}. The environment temperature was 22 - 27${ }^{\circ} \mathrm{C}$, and the humidity was kept in an interval of 50$\%$ - 70 $\%$. The categories of herbal medicines are: Astragalus, Liquorice, Chinese Angelica, Saposhnikovia
Divaricata, Radix Angelicae Pubescentis, Radix Angelicae Dahuricae, Notopterygium Incisum,
Codonopsis Pilosula, Radix Bupleuri, Ligusticum Chuanxiong Hort, Radix Peucedani, and Pueraria
Lobata. For each type of the medicines, 50 samples were collected with the following process:\\
\indent 1. Used an electrical pulverizer to grind the medicines into powders.\\
\indent 2. For each medicine sample (all-together 12x50=600 samples) , took 8 grams of its powder into a 125 ml glass container, and seal the container with a para-film.\\
\indent 3. Heated the sample powders in a thermostatic chamber (50 ${ }^{\circ} \mathrm{C}$) for 10 hours, and waited for another 10 hours to let the volatile gases diffuse in the glass container.\\
\indent 4. Took 10 mL gas mixtures in the headspace of each glass container as the objective gas, ready to be injected into sensors reaction chamber.

\subsection{Data processing and Feature Engineering}
The primitive data are the temporal signals from 16 sensors, in the form of voltage changes in 16 channels. First, the signals was calibrated by subtracting each sensors' baselines:
\begin{equation}
V=V_{S}-V_{0}
\end{equation}
Here $V_{s}$ denotes the recorded response of sensors, and $V_{0}$ denotes the baseline of sensors.\\
\indent According to the previous research, 8 commonly used features for each sensor were extracted, which turned out to be effective in pattern recognition with e-nose systems \cite{e-nose_public_dataset}. This feature extraction process output 128 features for a single sample:\\
\indent 1. The maximum value of V:
\begin{equation}
V_{\max }=\max (|V|)
\end{equation}
\indent 2. The area under the curve of V:
\begin{equation}
V_{i n t}=\int_{0}^{T} V(t) d t
\end{equation}
where T takes 340s, the same long as data recording period.\\
\indent 3-8. Exponential moving average of the derivative of V:
\begin{equation}
E_{a}(V)=[\min (y(k)), \max (y(k))], 2000<k<34,000
\end{equation}
where the discrete sampling exponential moving average and parameter $\alpha$ are set to be:
\begin{equation}
y(k)=(1-a) y(k-1)+a(V(k)-V(k-1))
\end{equation}
\begin{equation}
a=\frac{1}{100 * S R}, \frac{1}{10 * S R}, \frac{1}{S R}
\end{equation}
\begin{equation}
y(1)=a V(1) \text { , }
\end{equation}
where $SR$ represents the sampling rate (100 Hz in this study), and $E_{a}(V)$ denotes a vector of the smallest and largest values of $y(k)$. After performing the feature engineering, a feature matrix with size of (600 x 128) was extracted as our dataset, which constructed the sample space for 12 categories of herbal medicines in this study. 
\subsection{Study pipeline}
\begin{figure*}

    \centering
    \includegraphics[width=\linewidth]{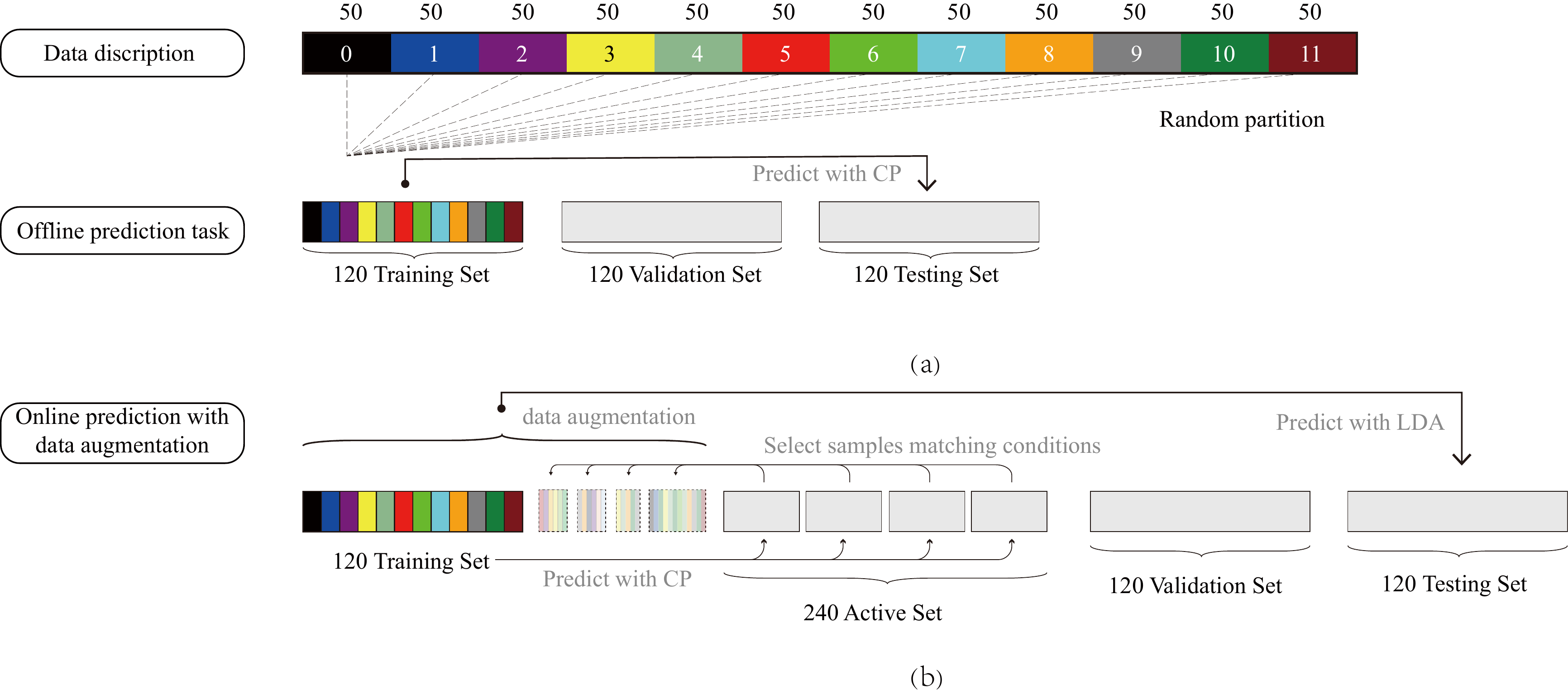}
    \caption{\textbf{The study pipeline of the two designed prediction tasks.}(a) The process of offline prediction: comparing the prediction accuracy of CPKNN and CPSC on testing set. (b) The process of online prediction with data augmentation: first employing CPKNN and CPSC to predict the active set and augment training set, and then comparing their effectiveness on improving prediction accuracy on testing set before/after data augmentation}
    \label{fig_study_pipeline}
\end{figure*}

\indent In this study, two tasks to compare the two CP frameworks (CPSC and CPKNN) were designed: 1) offline prediction and 2) online prediction with data augmentation, as shown in Fig. \ref{fig_study_pipeline}. The offline prediction task is the most prevalence machine learning application scenario. The dataset was randomly split into training set, validation set and testing set for 30 times, with a ratio of 4:1:1. First, training set was used to train the model (CPSC and CPKNN), and the hyperparameters were tuned according to the classification accuracy on the validation set. The hyperparameters included the shrink threshold $\Delta$ and temperature $T$ in CPSC, and the $K_{num}$ of kNN. In this offline task, three metrics were investigated to compare the CPSC and CPKNN performances: the classification accuracy on the testing set, the prediction credibility (as the major reliability information) given by the CPSC and CPKNN for the test predictions, and their runtime. Ideally, a CP framework should reach a high accuracy when the prediction credibility is high. Therefore, to compare the reliability of CPSC and CPKNN, the prediction accuracy with different intervals of credibility thresholds $\epsilon$ were calculated. 

Another application of the reliability given by conformal prediction is in the online prediction, where the data augmentation can be performed based on the reliability of the unlabeled data. In the online task, the dataset was partitioned into training set, active set (unlabelled data), validation set, and testing set, with a ratio of 1:2:1:1. The validation set is used for hyperparameter tuning, and the active set was the assumed unlabeled data available for online augmentation process. The online augmentation process follows the same pattern as our previous research \cite{augmentation}. As Fig. \ref{fig_study_pipeline} (b) shows, Linear Discriminant Analysis (LDA) \cite{LDA} is adopted as the basic classifier to predict testing set, as LDA generally performed better with the data augmentation process according to our previous research\cite{augmentation} and we aimed to only compare the influence of CPSC and CPKNN on data augmentation process. CPSC and CPKNN are applied in the data augmentation process. First, CPSC and CPKNN are trained with the initial training set, and then the 4 batches of active set were predicted and augmented into the training set in a batch-wise manner with these two CP frameworks. According to the previous research \cite{Neural_engineering_ICP}, in the augmentation process, only the sample satisfying two criteria can be chosen: the credibility of its label prediction (the largest P-value) should be 1) larger than the threshold $\epsilon$. 2) three times larger than the second-max P-value. These two criteria ensure that the predictions are with high credibility and that the prediction confidence is not too low. In this study, the effectiveness of data augmentation was evaluated by the improvement of LDA's classification accuracy on testing set. Furthermore, because CPSC and CPKNN output prediction regions which can be comprised of singleton predictions, empty predictions and multiple predictions, to observe the prediction dynamics of online prediction and compare the prediction characteristics of CPSC and CPKNN, the output of the region predictions were counted, and the following metrics were visualized: the accumulative counts of singleton predictions, the multiple predictions, the empty prediction, and the accepted samples in the augmentation process. Furthermore, to qualitatively compare the class balance in the augmentation process, we plotted the distribution of the predicted classes among the augmented data.

\subsection{Statistical tests}
To test the statistical significance of the performance difference between CPSC and CPKNN in two tasks (each task for 30 times), several statistical tests are leveraged: 1) Shapiro Wilk test for evaluating the normality of data, 2) Wilcoxon signed rank test or t-test for statistically comparing the prediction performance metrics (e.g. accuracy) from two groups, depending on the result of Shapiro Wilk test. While the Shapiro Wilk test rejected the normality assumption, Wilcoxon signed rank test is adopted rather than t-test.

\section{Result}

\subsection{Offline prediction}
\begin{figure}
    \centering
    \includegraphics[width=\linewidth]{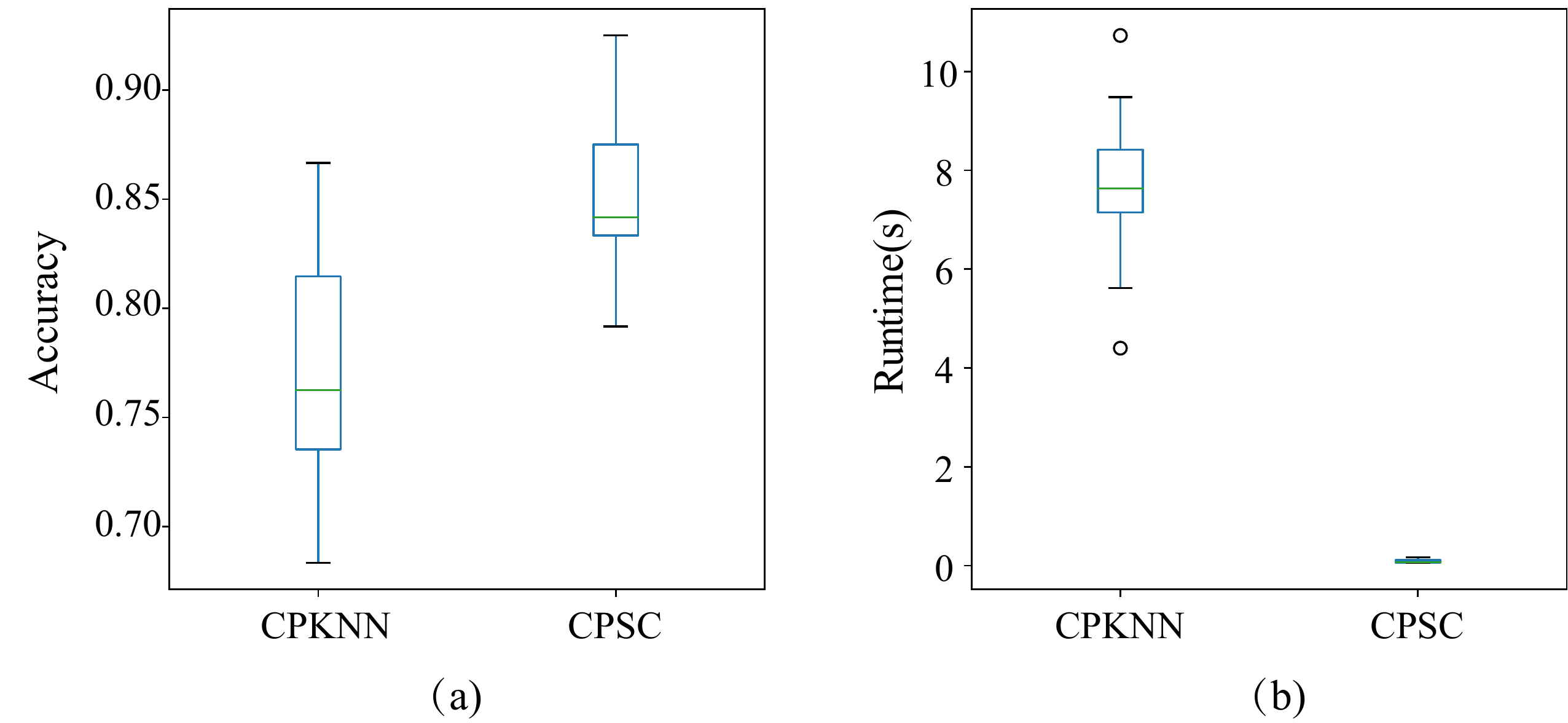}
    \caption{The offline prediction performance of CPSC and CPKNN, with 30 times of random dataset partition.(a) The box plots of prediction accuracy of CPSC and CPKNN. (b) The box plots of the time cost for training model and making predictions with CPSC and CPKNN.}
    \label{fig:offline_accuracy}
\end{figure}

 The accuracy of CPSC and CPKNN in offline prediction is illustrated in Fig. \ref{fig:offline_accuracy}, which shows that CPSC reaches a higher prediction accuracy (median: 0.854) than CPKNN (median: 0.763) with statistical significance (p $\leq$ 0.05) and runs much faster than CPKNN (p $\leq$ 0.05, and the computational cost analysis will be done in the last result subsection).
 
\begin{figure}
    \centering
    \includegraphics[width=\linewidth]{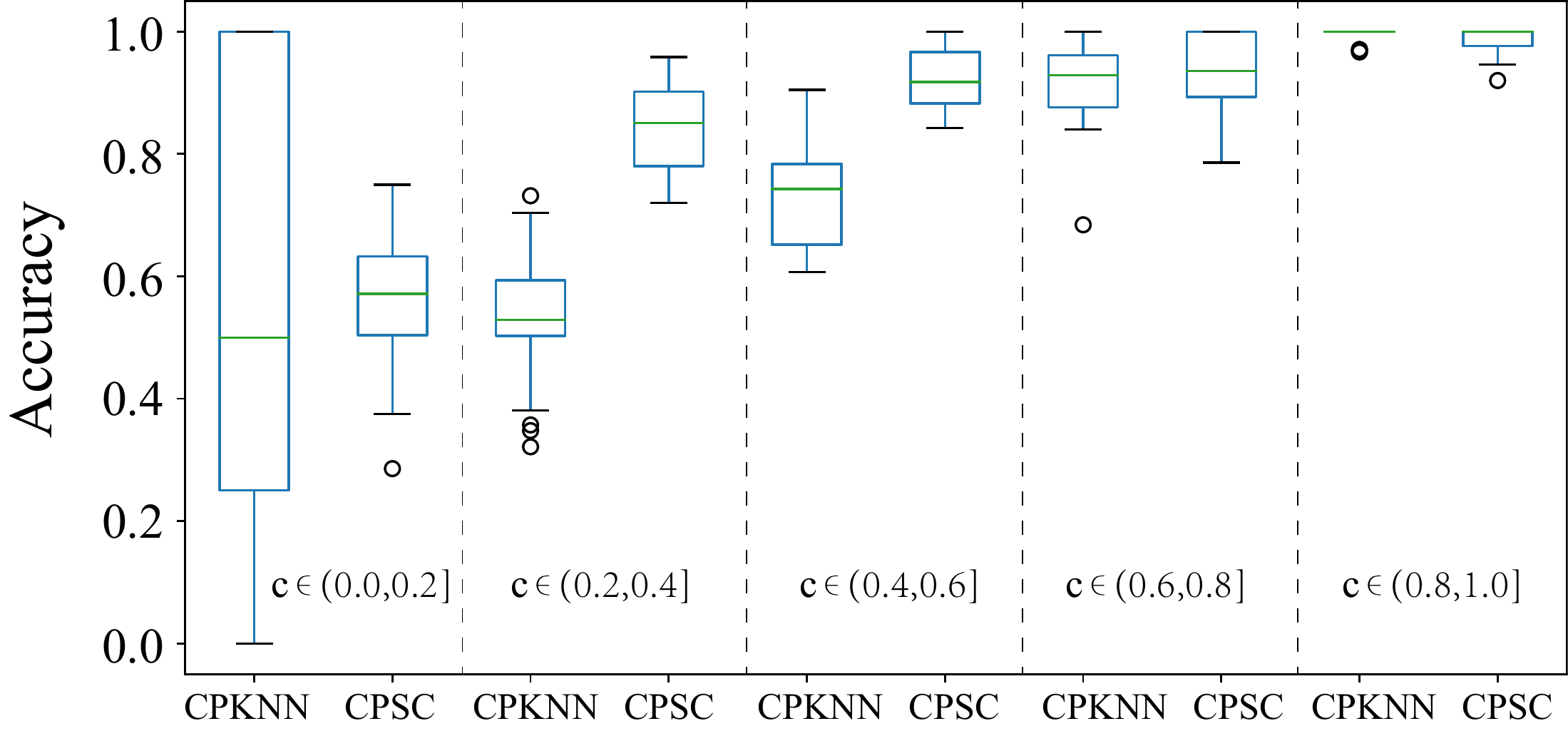}
    \caption{The offline prediction accuracy among the predictions within 5 credibility information intervals (each takes a span of 0.2), with 30 times of random dataset partition.}
    \label{fig:offline_comparison_interval}
\end{figure}

 As one important characteristic of CP is that this computational framework can provide the reliability of predictions \cite{cp_intro,1cp_introduction}, to verify the CP validity and test whether the predictions with high credibility are indeed more likely to be correct prediction, and to investigate how reliable the offline predictions CPKNN and CPSC make, a statistical comparison of the accuracy of the predictions with credibility values in different credibility intervals was done. According to the results shown in Fig. \ref{fig:offline_comparison_interval}, with a higher credibility ($c$) interval, the prediction accuracy of CPSC and CPKNN both increases. This indicates that 1) in CP framework, the credibility information $c$ is effective in assisting in decision making: if the prediction credibility is higher, the prediction is more likely to be true; 2) with the same credibility interval, predictions made by CPSC are more accurate than CPKNN, as CPSC generally achieves a higher prediction accuracy. These properties of CPSC enables the users to better leverage the credibility information to aid in decision making when compared with the conventional CPKNN.

\subsection{Online prediction with data augmentation}
\begin{figure}
    \centering
    \includegraphics[width=\linewidth]{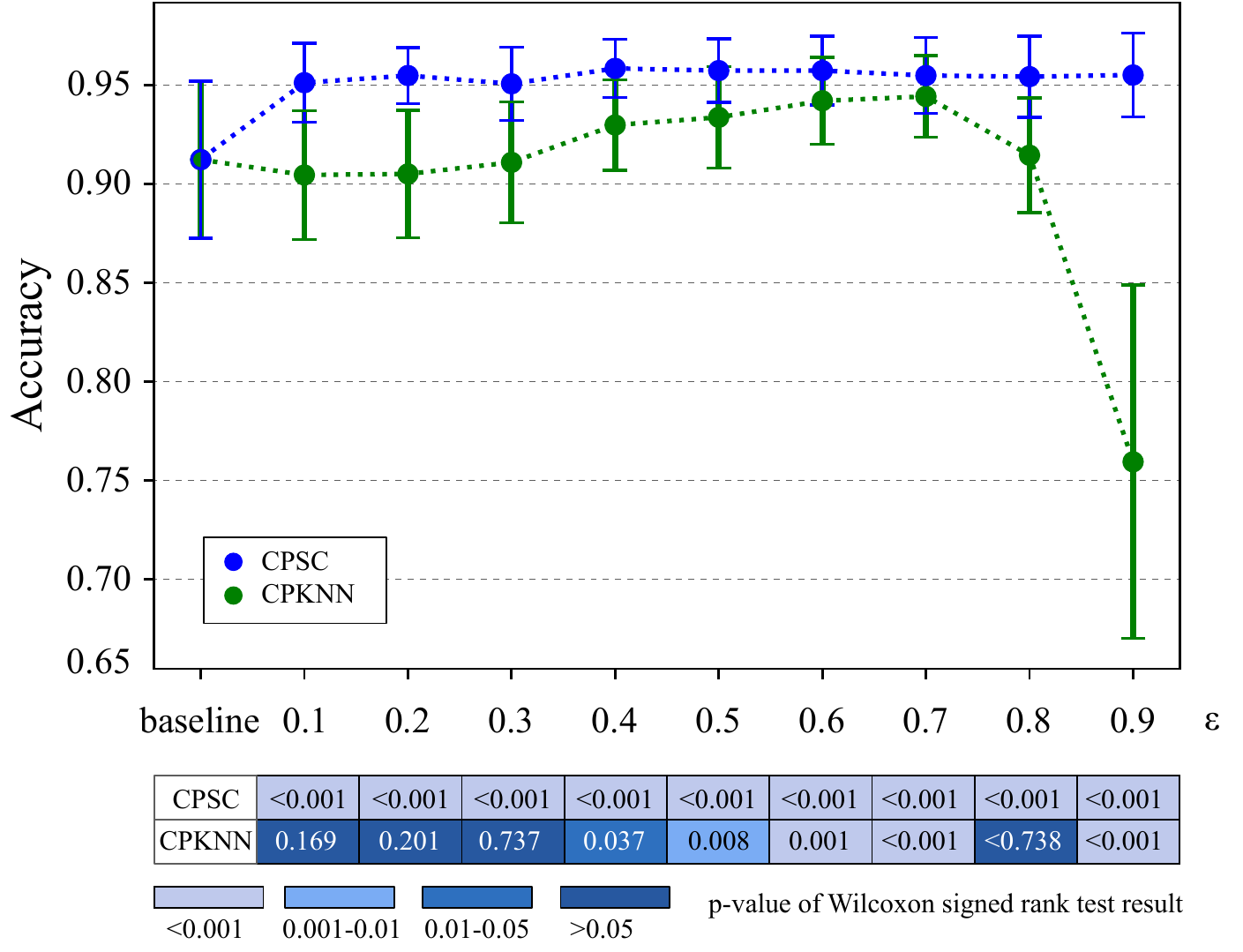}
    \caption{The online prediction accuracy on the testing set with data augmentation under a series of $\epsilon$ settings. The accuracy is calculated with LDA classifier after the entire data augmentation process has been done. Each group of accuracy results (with 30 times of random dataset partition) are compared with those given by the baseline (without augmentation). The statistical significance is tested by Wilcoxon signed-rank test.}
    \label{fig:accuracy_cpsc_cpknn}

\end{figure}

In addition to giving reliability information besides the predictions themselves for better decision making in offline scenarios, another major application of CP is for online prediction with data augmentation, where the augmented data can be filtered by the prediction reliability to improve the quality of augmented data \cite{4augmentation,3feature_engineering}. The accuracy of online prediction with data augmentation under different credibility thresholds $\epsilon$ (from 0.1-0.9) is displayed in Fig. \ref{fig:accuracy_cpsc_cpknn}. The result shows that CPSC robustly achieved a statistically significantly higher accuracy (p $\leq$ 0.05) after data augmentation with all $\epsilon$ (the credibility threshold that enables the prediction to be added into the training set) setting. Although both CPSC and CPKNN generally led to higher classification accuracy after data augmentation, however, CPKNN only showed accuracy improvement (p $\leq$ 0.05) in a subset of $\epsilon$ settings (0.4-0.7), and even showed statistically significantly decreases in accuracy when $\epsilon$ was set to be 0.9. The results shown in Fig. \ref{fig:accuracy_cpsc_cpknn} manifests that in a proper subset of $\epsilon$ settings (such as 0.2-0.7), a higher $\epsilon$ in data augmentation enables CPKNN to gain a lager accuracy improvement. However, when the threshold was set to be close to 1 (such as 0.8-0.9), CPKNN suffers an accuracy decline after data augmentation. In contrast, CPSC performed robustly with improved accuracy after data augmentation under all $\epsilon$ settings.  

\begin{figure}
    \centering
    \includegraphics[width=\linewidth]{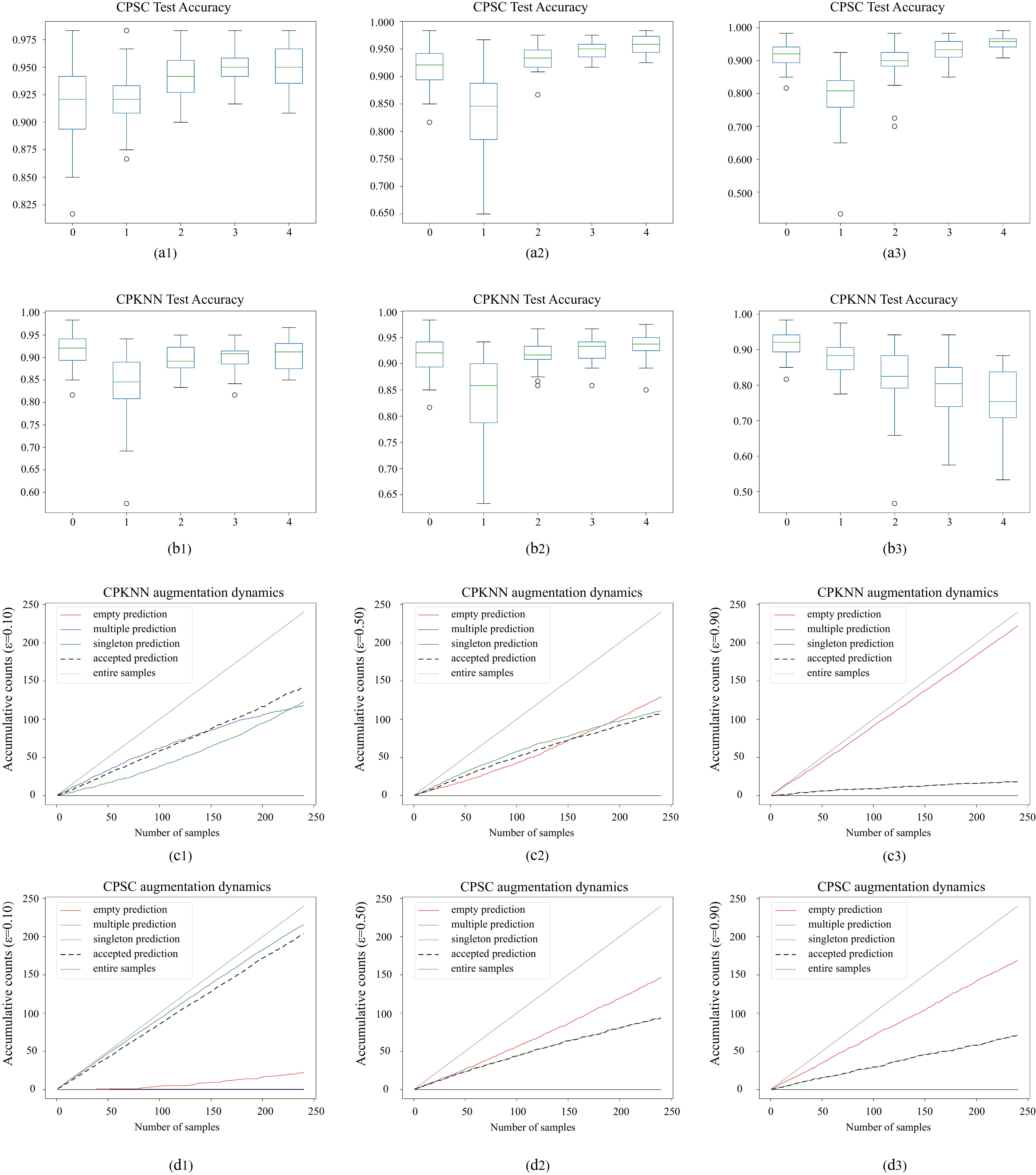}
    \caption{The prediction dynamics of online prediction with data augmentation. (a) and (b) respectively denotes the accuracy dynamics of CPSC and CPKNN. (c) and (d) denotes their accumulative counts of prediction indices. Index 1-3 denotes different $\epsilon$ settings: 0.1, 0.5, and 0.9.}
    \label{fig:4r3c}
\end{figure}

\begin{figure}
    \centering
    \includegraphics[width=0.7\linewidth]{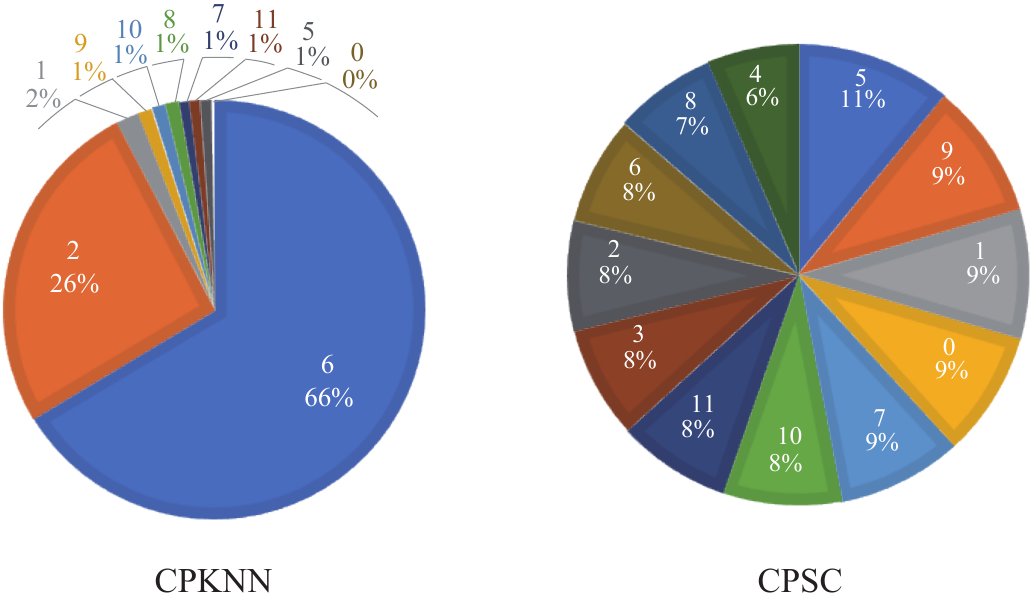}
    \caption{The distribution of the augmentation classes predicted by CPKNN and CPSC when $\epsilon$ is 0.9. }
    \label{fig_proportion}
\end{figure}

To better evaluate the prediction types given by CPSC and CPKNN, the prediction dynamics of online augmentation process is illustrated in Fig.\ref{fig:4r3c}, which presents the classification accuracy after each batch of data augmentation in Fig.\ref{fig:4r3c} (a) and (b), and the accumulative counts of specific types of predictions: accepted prediction (the predictions which satisfy the reliability requirements in Section 2.7), singleton predictions (the predictions with only one predicted label in the region predictions), multiple predictions (the predictions with more than one predicted labels in the region predictions), and empty predictions (the predictions with no label in the region predictions) in Fig. \ref{fig:4r3c} (c) and (d). Here, three $\epsilon$ are taken as examples: 0.1, 0.5, and 0.9, and the results are displayed in (Fig. S1), including the result of $\epsilon$ from 0.1, 0.3, 0.5, 0.7 and 0.9. For the prediction accuracy, in cases where the final accuracy was improved (CPSC ((a1)-(a3)) and CPKNN cases ((b1)-(b2))), the accuracy dynamics manifest a similar pattern: when the first batch was augmented, the accuracy slightly declined, possibly because the models did not learn a robust relationship based on the initial training data. When more active data were added into the training set via augmentation, the accuracy increased to the initial level and finally outperformed the baseline (p $\leq$ 0.05). However, in (b3) (CPKNN with $\epsilon$ 0.9), the accuracy was gradually declining with the addition of every batch of active data. Based on the accumulative counts of various types of predictions, when $\epsilon$ was fixed, CPKNN generally output much more multiple predictions (i.e. more than one predicted labels in the prediction region) while CPSC generally output more singleton predictions. As for the empty predictions, CPSC output more empty predictions when $\epsilon$ was lower than 0.5, but CPKNN gave more of them when $\epsilon$ is higher than 0.5. As for the accepted predictions, CPSC generally took in more augmented samples than CPKNN, especially when $\epsilon$ comes to 0.9. To 
better investigate the augmentation process when $\epsilon$ was 0.9, the distribution of the predicted classes of augmented samples from CPSC and CPKNN was illustrated in Fig. \ref{fig_proportion}. In general, CPSC tends to be more certain in making predictions: the predicted labels are with high credibility values. Therefore, its prediction region $\Gamma^{\varepsilon}$ always includes only a singleton prediction, and with the same credibility threshold $\epsilon$, CPSC generally accepts more samples than CPKNN does. Besides, compared with CPKNN, CPSC kept a more balanced distribution in predicted class when augmenting data.

\subsection{Computational cost}
Fig.~\ref{fig:offline_accuracy} (b) displays the box plot of runtime for training model and classifying 120 cases in offline task with  CPKNN and our proposed CPSC, where the median runtime of CPKNN and CPSC over 30 trials are about $7.5$s and $0.05$s, respectively. We used a personal computer ( Intel(R) Core(TM) i7-8550U CPU@1.80GHz 2.00 GHz, and a RAM with 16.0 GB) to train the models. It is obvious that, compared with CPKNN, the computational cost of CPSC decreases considerably. Furthermore, the runtime of CPKNN ranges over $5.5$-$9.5$s, which is clearly larger than the runtime variance in terms of the CPSC.

Theoretically, the computational cost of CPKNN is dominated by the calculation of nonconformaity measurement $\alpha_i$. Based on the \eqref{Nonconformity measurement CPKNN}, the computational complexity of CPKNN can be stated as $O(pn)$, where $p$ denotes the overall feature number of collected samples, $n$ represents the size of training dataset. On the other hand, the computational time of CPSC mainly depends on the standard deviation update and the distance computation between new sample with , which are indicated as to \eqref{eq:std_update}, \eqref{eq:new_cluster}, respectively. And the total computational complexity of CPSC is supposed to be $O(n+pK)$, where $K$ denotes the class number. Normally, the size of training dataset ($n$) surpasses the number of overall categories ($K$) extremely. Therefore, training CPKNN model consumes several times as long as the CPSC model.

\section{Discussion}

In this study, CPSC is proposed as a novel conformal predictor, and it is compared against the conventional CP framework CPKNN which has already been proved to be effective in many classification tasks for reliability quantification and data augmentation, in such application fields as herbal medicine classification \cite{E_nose_zhan2018sensors,Diao_dendrobium,E_nose_WZ_cpknn,augmentation} and lung cancer detection\cite{2cpknn_lung_cancer}. This study investigates CPSC and CPKNN in classifying 12 categories of herbal medicines with the e-nose system, where two tasks are designed: offline prediction (to compare the accuracy and verify the ability of reliability quantification), and online prediction with data augmentation (to compare the effectiveness of data augmentation with CPSC/CPKNN in improving classificaiton accuracy). In these two tasks, several metrics are studied: the accuracy and reliability of offline prediction, the effectiveness of data augmentation in online prediction, and the computational cost. In task 1 (offline prediction), CPSC achieves a higher accuracy with a much faster speed with statistical significance. In task 2 (online prediction with data augmentation), CPSC stably improves the prediction accuracy with a series of credibility thresholds. 
Furthermore, the augmentation effect tends to be better with the increase of $\epsilon$, as the threshold of filtering the predictions is getting stricter. However, CPKNN suffers a decline when $\epsilon$ reaches 0.9. According to Fig. \ref{fig_proportion}, most of the augmented samples filtered by CPKNN belong to class 6 or 2, which shows an unbalanced distribution of the augmentation classes predicted by CPKNN. This may be because of the more evident feature dissimilarities of class 6 from other classes, which is also shown in the heatmap in our previous publications \cite{augmentation}. In contrast, CPSC filters the augmented samples in a much more balanced manner with more cases in diverse classes, which may finally lead to the better performances after data augmentation with CPSC.

 As for the reasons why CPSC generally outperforms CPKNN in both the offline and online prediction tasks, we believe that it may be attributed to the fact that CPSC attenuates the negative influences of relatively irrelevant features via the shrunken centroids, and the fact that CPSC better summarizes the classification information with the representative centroids for each class. Therefore, the variance and bias caused by curse of dimensionality could be alleviated as the feature space is more compact. The reason for CPSC's advantage on the computational cost is that CPSC 1) attenuates the irrelevant features thus reduces dimensionality ; 2) makes predictions based on the distance with regularized class centroids. Therefore, the times of Euclidean distance calculation for a test sample is proportional to the number of centroids. On the contrary, CPKNN calculates and sorts the Euclidean distance between a test sample and each training sample in a pair-wise manner. The optimization in Euclidean distance calculation (both the shrunken feature space and reduced computational cost) enables CPSC to perform faster, especially when dealing with high dimensional data. Furthermore, the parameters (temperature and shrinkage) in CPSC help users with the freedom to adjust the regularization strength of prediction and calibrate the reliability given by CPSC with expert knowledge. To be exact, the newly proposed parameter $T$ in CPSC makes the probabilities distribution among class more balanced, and the shrinking threshold $\delta$ controls the extent of dimensionality reduction and noise attenuation. When these parameters were set in a proper interval, they have little impact on prediction accuracy but may optimize the reliability of the predictions where the users can calibrate the reliability quantified by CPSC with their field experience and expertise.
 
The CPSC proposed by us is one component in the broad category of conformal prediction (CP) algorithms, which is featured with the ability to provide the reliability information related to the classification prediction \cite{1cp_introduction} in a distribution-free manner. This important characteristic can help users to make a decision with additional consideration of reliability and prediction confidence, which can also be referred to as the prediction mindset of the classifier. For example, over the past years, machine learning and deep learning have shown their great potential in assisting in medical diagnosis, such as the COVID-19 patient outcome prediction, lung cancer diagnosis and traumatic brain injury monitoring \cite{E_nose_zhan2018sensors,brain_strain_zhan,2cpknn_lung_cancer}. However, an incorrect decision made by machine learning models in medical diagnosis might result in disastrous consequence. For example, in cancer detection, a patient receiving a false negative diagnosis result might miss the best period for treatment. On the contrary, a false positive diagnosis result might lead the patient to suffer from unnecessary chemotherapy, which does harm to the patient's normal tissues. Therefore, to improve the reliability of machine learning applications in medical diagnosis, the credibility, as one type of the reliability information provided by CP, can be used to facilitate more reliable medical diagnosis: a high prediction credibility indicates that the results are generally more reliable for the user to trust. However, a low credibility informs that the prediction may not be reliable enough so that the doctors may need to consider other diagnostic approaches to confirm the decision output by the machine learning system, instead of drawing an arbitrary conclusion. In addition to ensure the reliability in decision, CP has manifested its effectiveness in online prediction with data augmentation. With proper thresholds on credibility, CP filters and predicts the unlabeled samples that have high conformity with initial training set. In this way, the CP leverages the reliability as the "prediction mindset" and utilizes it to filter the predictions for data augmentation with higher quality. As a result, with the CP and the reliability thresholding, the classifiers can learn to be more powerful and accurate with the increasing number of samples.

Although CP has lots of unique benefits which are mentioned above, it still has something worth attention from the users to ensure online prediction with data augmentation. First, the threshold $\epsilon$ should stay in a proper interval to balance the quality and quantity of the augmented data. As Fig. \ref{fig:accuracy_cpsc_cpknn} and Fig. \ref{fig:4r3c} show, when $\epsilon$ is above 0.7, the quantity of augmented samples declines and the data augmentation can not always ensure a statistically significant improvement on prediction accuracy. This is because the higher $\epsilon$ is, although the samples filtered by a stricter criteria can be more reliable (i.e., the predictions with higher credibility values are with higher prediction accuracy, as shown in Fig. \ref{fig:offline_comparison_interval}), the fewer samples can be added in the training set, and the distribution of the augmentation classes is more likely to be unbalanced (which is shown in using CPKNN in Fig. \ref{fig_proportion}, where the majority of the augmented samples have the predicted class as 6 or 2) as CP more readily picks up the easy-to-predict classes. In this case, the augmentation with high $\epsilon$ might have directly led the model to overfit the training data. However, a low $\epsilon$ might further regularize the model by allowing many less reliable augmented samples to be added to the training set. We suppose the reason why more augmented samples matter in this case is: the initial training set may not enable the models to learn a robust pattern for each class and a robust decision boundary between each pair of classes, while the augmented samples aid in the expansion of the feature space covered by the training samples and add in more information related to the data distribution. As a result, the larger quantities of unlabelled data may facilitate the modeling process. Therefore, the balance between augmentation's quality and quantity is important, which requires a proper $\epsilon$ setting. Second, the parameters ( $K_{num}$ in CPKNN, $T$ and $\delta$ in CPSC) should be tuned based on the need of users, which has been illustrated in the previous discussion paragraphs. 

Although the novel CPSC proposed in this paper outperformed the conventional CPKNN in many aspects in the offline and online tasks, there are still limitations in our research which must be mentioned. First, there lacks a comparison between CPSC and other more complicated CP framework, such as CPSVM and CP-LSTM \cite{Neuro_computing}, which may be able to perform better than CPKNN. In this study, we focused on the comparison with CPKNN. This is because CPKNN is the first proposed CP framework, which has been proved effective by many datasets. Therefore, it can be regarded as a proper baseline to be compared with. Another reason is that we presume that CPSVM and CP-LSTM are more flexible models driven by the data which take even more time on computation when compared with CPKNN due to the optimization problem setting. Second, for each dataset, only one type of feature engineering method is adopted. We chose the feature engineering approahces which have shown effective or the most effective in terms of classification accuracy in the previous publications \cite{e-nose_public_dataset,E_nose_zhan2018sensors}. However, we believe that CPSC can display even more of its advantages when facing high dimensional dataset if we change the feature sets, such as using down-sample and FFT features \cite{feature_engineering_cac} (the number of features is more than 318). Therefore, as for future work, CPSC will be compared with other types of CP framework and applied on dataset with even higher dimensionality to further validate the model's advantage.   

\section{Acknowledgement}
L. Liu and X. Zhan contributed equally to this work. The work is supported by the Natural Science Foundation of China (Grant No. 61773342) and the Science Fund for Creative Research Groups of NSFC (Grant No.61621002). 
\newpage
\section{Supporting Information}
Supplementary matrials are as follows:

\setcounter{figure}{0}
\renewcommand{\thefigure}{S\arabic{figure}}


\begin{figure}[htbp]
\centering
\centering
\includegraphics[scale=0.1]{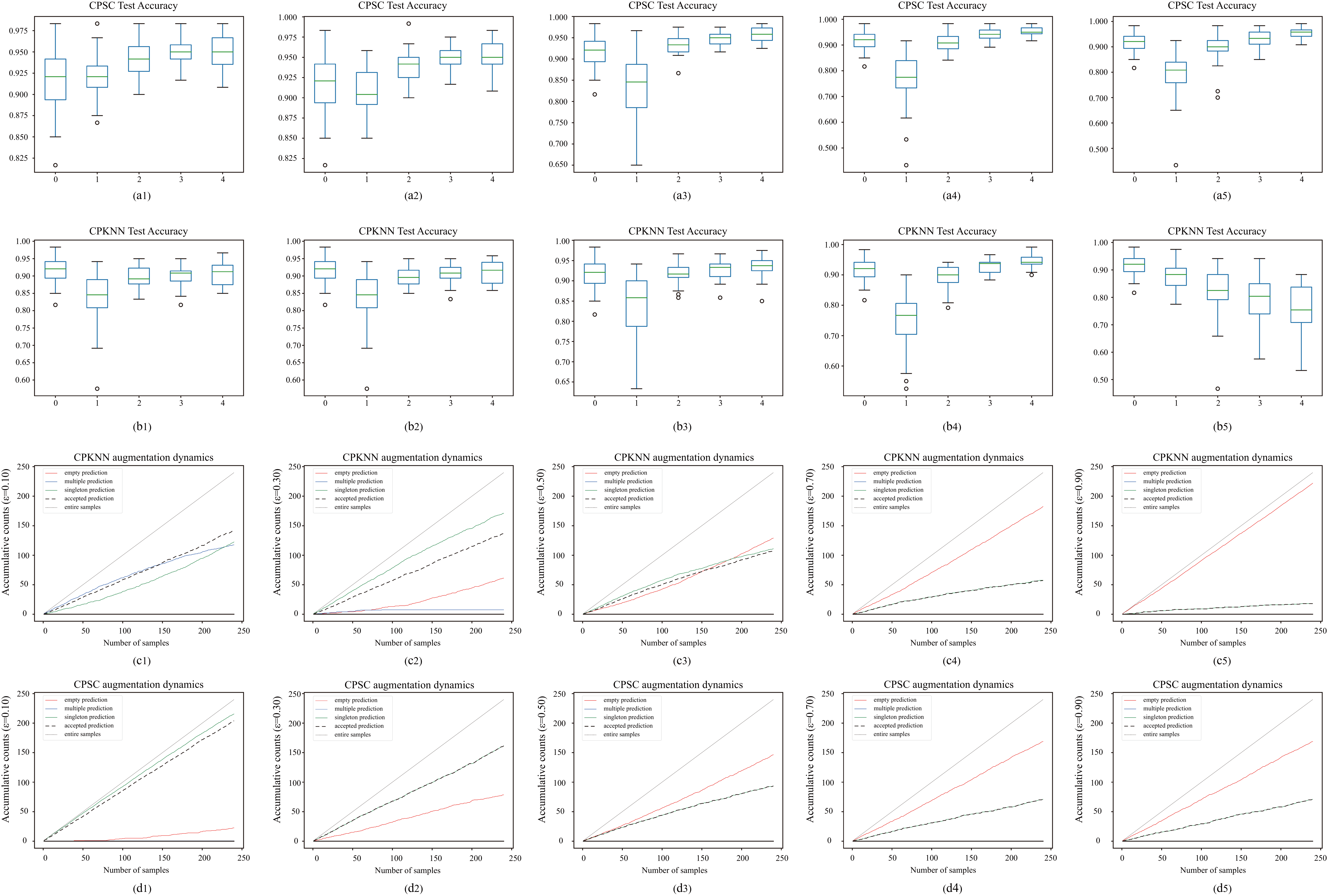}
\caption{\textbf{The prediction dynamics of online prediction with data augmentation. (a) and (b) respectively denotes the accuracy dynamics of CPSC and CPKNN.} (c) and (d) denotes their accumulative counts of prediction indices. Index 1-5 denotes different $\epsilon$ settings: 0.1, 0.3, 0.5, 0.7 and 0.9. }
\end{figure}

\newpage
\clearpage
\section{Acknowledgments}

	The work is supported by the Natural Science Foundation of China (Grant No. 61773342) and the Science Fund for Creative Research Groups of NSFC (Grant No.61621002). In this study, the data was collected at Zhejiang University. The authors appreciate the research equipment and assistance from the State Key Laboratory of Industrial Control Technology, Institute of Cyber-Systems and Control, Zhejiang University.


\section{Conflict of interests}
The authors declare no conflict of interests.

\nolinenumbers


\newpage
 \clearpage
\bibliography{ref}
\bibliographystyle{IEEEtran}


\end{document}